\documentclass[12pt]{article}
\usepackage{graphicx}

\newcommand{\beq}{\begin{equation}}
\newcommand{\eeq}{\end{equation}}
\newcommand{\beqa}{\begin{eqnarray}}
\newcommand{\eeqa}{\end{eqnarray}}
\newcommand{\non}{\nonumber}
\newcommand{\lab}{\label}
\newcommand{\bra}{\langle}
\newcommand{\ket}{\rangle}

\begin{document}

\title{Another convex combination of product states for the separable Werner state}

\author{Hiroo Azuma${}^{1,}$\thanks{On leave from Canon Inc., 5-1,
Morinosato-Wakamiya, Atsugi-shi, Kanagawa, 243-0193, Japan.
E-mail: h-azuma@lab.tamagawa.ac.jp}
\ \ 
and
\ \ 
Masashi Ban${}^{2,3,}$\thanks{E-mail: m-ban@harl.hitachi.co.jp}
\\
\\
{\small ${}^{1}$Research Center for Quantum Information Science,}\\
{\small Tamagawa University Research Institute,}\\
{\small 6-1-1 Tamagawa-Gakuen, Machida-shi, Tokyo 194-8610, Japan}
\\
{\small ${}^{2}$Advanced Research Laboratory, Hitachi Ltd.,}\\
{\small 2520 Akanuma, Hatoyama, Saitama 350-0395, Japan}
\\
{\small ${}^{3}$CREST, Japan Science and Technology Agency,}\\
{\small 1-1-9 Yaesu, Chuo-ku, Tokyo 103-0028, Japan}
}

\date{February 10, 2006}

\maketitle

\begin{abstract}
In this paper, we write down the separable Werner state
in a two-qubit system explicitly
as a convex combination of product states,
which is different from the convex combination
obtained by Wootters' method.
The Werner state in a two-qubit system
has a single real parameter
and
varies from inseparable state to separable state
according to the value of its parameter.
We derive a hidden variable model that is induced
by our decomposed form for the separable Werner state.
From our explicit form of the convex combination of product states,
we understand the following:
The critical point of the parameter for separability of the Werner state
comes from positivity of local density operators of the qubits.
\end{abstract}

\section{Introduction}
\lab{section-introduction}
The Einstein-Podolsky-Rosen paradox and Bell's pioneering works
reveal that no hidden variable model can reproduce all predictions of quantum mechanics
\cite{Einstein-Podolsky-Rosen,Bell,Clauser-Horne-Shimony-Holt}.
Thus, quantum correlation is essentially different from classical correlation.
Motivation of quantum information theory,
which many researchers have been eager to study for the last several decades,
is to obtain a deep understanding of the quantum correlation.

A bipartite quantum system is separable
if its density matrix can be written
as a convex combination of product states.
A separable quantum system always admits the hidden variable interpretation.
However, the converse is not necessary true.
Werner constructs a family of bipartite states,
which are characterized by a single real parameter.
He shows some inseparable states that belong to this family
admit the hidden variable interpretation
\cite{Werner}.
The states of this family are called the Werner states.
Moreover, Popescu indicates that
the inseparable Werner states admitting hidden variable models
reveal nonlocal correlation under a sequence of measurements,
where the second measurement depends on an output of the first measurement
\cite{Popescu}.

A criterion of separability for a two-qubit system
is conjectured by Peres and established by Horodecki {\it et al}.
\cite{Peres,Horodecki-Horodecki-Horodecki}.
In a two-qubit system,
the Werner state has a single real parameter
and varies from inseparable state to separable state
according to the value of its parameter.
Using Peres-Horodeckis' criterion,
we can fix the critical point of the parameter
between the separable and inseparable states.

The Werner state is finding wide application in the quantum information processing.
It often appears as an intermediate
during the quantum purification protocol
\cite{Bennett-DiVincenzo-Smolin-Wootters,Murao-Plenio-Popescu-Vedral-Knight}.
Thus, we can expect that the Werner state plays an important role
in the process of local quantum operations and classical communications (LQCC).
Various properties of the Werner state under LQCC is
investigated by Hiroshima and Ishizaka
\cite{Hiroshima-Ishizaka}.

As mentioned above,
the Werner state has many interesting properties.
However, we do not know
why the Werner state in a two-qubit system
changes from inseparable state to separable state suddenly
at the critical point of its parameter.
To examine the physical meaning of the critical point,
we have to know an explicit form of a convex combination of product states
for the separable Werner state.
In this paper, we investigate a convex combination of product states
for the separable Werner state,
which is different from the convex combination obtained by Wootters' method
~\cite{Wootters}.
(A convex combination of product states for a given separable density matrix
is not unique generally.)

The decomposition obtained by Wootters' method is an ensemble of four pure states.
By contrast,
our decomposition is an integral of a product state
with a probability distribution function over a continuous variable.
Our decomposed form seems simpler than the decomposed form
obtained by Wootters' method,
and thus it may give us some insight.
This is an advantage of our result.
Looking at our decomposed form,
we can understand that the critical point of the parameter
for separability of the Werner state
comes from positivity of local density operators of the qubits.
Furthermore, our result produces a hidden variable model
because the convex combination of product states
always causes the hidden variable interpretation.

Here, we give a brief summary of Wootters' results in Ref.~\cite{Wootters}.
Wootters shows an explicit formula for the entanglement of formation
of an arbitrary two-qubit system as a function of its density matrix.
We can judge whether or not a given two-qubit density matrix is entangled
by a value of its entanglement of formation.
The density matrix is not entangled if its entanglement of formation is
equal to zero,
and
the density matrix is entangled if its entanglement of formation is
more than zero.
Thus, we can use Wootters' formula instead of Peres-Horodeckis' criterion.

Wootters also shows how to construct an entanglement-minimizing decomposition
of an arbitrary two-qubit density matrix.
In this decomposition, the density matrix is described by a convex combination of pure states,
and the average entanglement of the pure states is equal to the entanglement of formation.
Thus, if we decompose a separable two-qubit density matrix
according to Wootters' method,
we obtain an ensemble of pure states,
each of which has no entanglement.
Hence, in general,
Wootters' decomposition gives us a convex combination of product states
for a separable density matrix explicitly.
In Appendix~\ref{appendix-wootters-decomposition},
we write down the decomposition obtained by Wootters' method
for the separable Werner state.

In the rest of this section,
we introduce the Werner state for a two-qubit system
and
examine its separability by Peres-Horodeckis' criterion.
In Sec.~\ref{section-Werner-state-and-hidden-variable-interpretation},
we investigate the relation
between the separable Werner state and the hidden variable interpretation.
In Sec.~\ref{section-derivation-hidden-variable-model-Werner-state},
we derive the explicit form of the convex combination of product states
for the separable Werner state.
In Sec.~\ref{section-discussions} we give a brief discussion.
In Appendix~\ref{appendix-wootters-decomposition},
we describe the decomposition of the separable Werner state
obtained by Wootters' method.

The Werner state is given by the following density operator
on a four-dimensional Hilbert space
$\mathcal{H}_{\mbox{\scriptsize A}}
\otimes
\mathcal{H}_{\mbox{\scriptsize B}}$
spanned by two qubits A and B:
\beq
W(q)
=
q|\Psi^{-}\ket\bra\Psi^{-}|
+
\frac{1-q}{4}\mbox{\boldmath $I$}_{(4)},
\lab{definition-Werner-state}
\eeq
where $0\leq q \leq 1$.
$\mbox{\boldmath $I$}_{(4)}$ is the identity operator on
$\mathcal{H}_{\mbox{\scriptsize A}}
\otimes
\mathcal{H}_{\mbox{\scriptsize B}}$.
$|\Psi^{-}\ket$ is one of the Bell states that are maximally entangled
on the two-qubit system
and it is given by
\beq
|\Psi^{-}\ket
=
\frac{1}{\sqrt{2}}
(
|0\ket_{\mbox{\scriptsize A}}|1\ket_{\mbox{\scriptsize B}}
-
|1\ket_{\mbox{\scriptsize A}}|0\ket_{\mbox{\scriptsize B}}
).
\eeq
$W(q)$ defined by Eq.~(\ref{definition-Werner-state}) satisfies
\beq
W(q)^{\dagger}=W(q),
\eeq
\beq
\mbox{Tr}W(q)=1,
\eeq
\beq
\bra\psi|W(q)|\psi\ket\geq 0 \quad\quad\forall|\psi\ket.
\eeq
Because of the above properties,
we can regard $W(q)$ as a density operator.

We can judge whether $W(q)$ is separable or inseparable,
that is,
whether $W(q)$ is disentangled or entangled,
from Peres-Horodeckis' criterion.
According to Peres-Horodeckis' criterion,
defining the partial transposition of $W(q)$ as $\tilde{W}(q)$,
$W(q)$ is separable if all eigenvalues of $\tilde{W}(q)$ are non-negative,
and
$W(q)$ is inseparable if one of eigenvalues of $\tilde{W}(q)$ is negative.
Let us examine eigenvalues of $\tilde{W}(q)$ below.

First of all, we give a matrix representation of $W(q)$ in a ket basis
$\{|i\ket_{\mbox{\scriptsize A}}|j\ket_{\mbox{\scriptsize B}}:
i,j\in\{0,1\}\}$
as follows:
\beq
W(q)
=
\frac{1}{4}
\left(
\begin{array}{cccc}
1-q & 0   & 0   & 0   \\
0   & 1+q & -2q & 0   \\
0   & -2q & 1+q & 0   \\
0   & 0   & 0   & 1-q
\end{array}
\right).
\lab{matrix-representation-Wq}
\eeq
Thus, we obtain a matrix representation of $\tilde{W}(q)$
as follows:
\beq
\tilde{W}(q)
=
\frac{1}{4}
\left(
\begin{array}{cccc}
1-q & 0   & 0   & -2q \\
0   & 1+q & 0   & 0   \\
0   & 0   & 1+q & 0   \\
-2q & 0   & 0   & 1-q
\end{array}
\right).
\lab{partial-transpose-Wq}
\eeq
In Eq.~(\ref{partial-transpose-Wq}),
the density operator is subjected to transposition
on the Hilbert space $\mathcal{H}_{\mbox{\scriptsize B}}$
spanned by the qubit B.
By some calculation,
we obtain three-fold degenerate eigenvalues,
$(1+q)/4$,
and the last eigenvalue, $(1-3q)/4$,
for $\tilde{W}(q)$.
Hence, $W(q)$ is separable for $0\leq q\leq 1/3$
and inseparable for $1/3<q\leq 1$.

\section{The separable Werner state and the hidden variable interpretation}
\lab{section-Werner-state-and-hidden-variable-interpretation}
From the discussion given in the previous section,
we find that $W(q)$ is separable for $0\leq q\leq 1/3$.
The separable $W(q)$ has to be rewritten as the convex combination of product states
and therefore it admits the hidden variable interpretation.
In this section, we investigate relation
between the separability of $W(q)$ and the hidden variable interpretation.

In general, we can rewrite the separable $W(q)$ in the form:
\beq
W(q)
=
\sum_{\lambda}
p_{\lambda}
(\rho_{\mbox{\scriptsize A},\lambda}
\otimes
\rho_{\mbox{\scriptsize B},\lambda}),
\lab{W(q)-product-state-representation}
\eeq
where
$0\leq p_{\lambda}\leq 1$,
$\sum_{\lambda}p_{\lambda}=1$,
and $\rho_{\mbox{\scriptsize A},\lambda}$ and $\rho_{\mbox{\scriptsize B},\lambda}$
represent density operators of the qubits A and B, respectively.
Here, we may regard the index $\lambda$ in Eq.~(\ref{W(q)-product-state-representation})
as a continuous variable.
Moreover, we can describe an arbitrary one-qubit density operator $\rho$ as
\beq
\rho
=
\frac{1}{2}
(\mbox{\boldmath $I$}_{(2)}
+
\mbox{\boldmath $a$}\cdot\mbox{\boldmath $\sigma$}),
\eeq
where
$\mbox{\boldmath $I$}_{(2)}$ represents the identity operator
on a two-dimensional Hilbert space spanned by a single qubit,
and
$\mbox{\boldmath $a$}$ represents an arbitrary three-dimensional real vector
whose norm is equal to or less than unity.
$\mbox{\boldmath $\sigma$}$ stands for a three-dimensional vector
whose three components are Pauli matrices,
$\mbox{\boldmath $\sigma$}=(\sigma_{x},\sigma_{y},\sigma_{z})$.

From the above consideration,
we can rewrite $W(q)$ defined in Eq.~(\ref{W(q)-product-state-representation})
as follows:
\beqa
W(q)
&=&
\int
d\lambda\;
p(\lambda)
(\rho_{\mbox{\scriptsize A}}(\lambda)
\otimes
\rho_{\mbox{\scriptsize B}}(\lambda)) \non \\
&=&
\int
d\lambda\;
p(\lambda)
\frac{1}{2}
(\mbox{\boldmath $I$}_{(2)}
+
\mbox{\boldmath $a$}(\lambda)\cdot\mbox{\boldmath $\sigma$})_{\mbox{\scriptsize A}}
\otimes
\frac{1}{2}
(\mbox{\boldmath $I$}_{(2)}
+
\mbox{\boldmath $b$}(\lambda)\cdot\mbox{\boldmath $\sigma$})_{\mbox{\scriptsize B}},
\lab{W(q)-product-state-representation-continuous-lambda}
\eeqa
where
\beq
\int
d\lambda\;
p(\lambda)
=
1,
\lab{probability-preserving-condition}
\eeq
and
\beq
|\mbox{\boldmath $a$}(\lambda)|\leq 1,
\quad\quad
|\mbox{\boldmath $b$}(\lambda)|\leq 1.
\lab{norm-condition}
\eeq
Although we write $\lambda$ as a single variable
in Eq.~(\ref{W(q)-product-state-representation-continuous-lambda}),
we can consider that $\lambda$ stands for multiple variables.
Moreover,
because
$\lambda$, $p(\lambda)$, $\mbox{\boldmath $a$}(\lambda)$, and $\mbox{\boldmath $b$}(\lambda)$
depend on $q$,
strictly speaking,
we have to write them as
$\lambda(q)$,
$p(q,\lambda(q))$,
$\mbox{\boldmath $a$}(q,\lambda(q))$,
and
$\mbox{\boldmath $b$}(q,\lambda(q))$.
However,
for simplicity,
we omit $q$ from their notations.
[We never insist that
$\lambda$, $p(\lambda)$, $\mbox{\boldmath $a$}(\lambda)$, and $\mbox{\boldmath $b$}(\lambda)$
do not depend on $q$.]
Furthermore,
we have to pay attention to the fact that
the convex combination of product states for $W(q)$
given in Eq.~(\ref{W(q)-product-state-representation-continuous-lambda})
is not unique.

The convex combination of product states for $W(q)$
given in Eq.~(\ref{W(q)-product-state-representation-continuous-lambda})
admits the hidden variable interpretation.
We can understand this fact from the following explanation.
Let us perform orthogonal measurements
by Hermitian operators,
$E(\mbox{\boldmath $l$})_{\mbox{\scriptsize A}}$ and
$E(\mbox{\boldmath $m$})_{\mbox{\scriptsize B}}$,
on the qubits A and B, respectively.
We assume that
$E(\mbox{\boldmath $l$})_{\mbox{\scriptsize A}}$ and
$E(\mbox{\boldmath $m$})_{\mbox{\scriptsize B}}$
are given by the following form:
\beq
\left\{
\begin{array}{lll}
E(\mbox{\boldmath $l$})_{\mbox{\scriptsize A}}
=\mbox{\boldmath $l$}\cdot\mbox{\boldmath $\sigma$}_{\mbox{\scriptsize A}}
&\quad&
\mbox{for $|\mbox{\boldmath $l$}|=1$}, \\
E(\mbox{\boldmath $m$})_{\mbox{\scriptsize B}}
=\mbox{\boldmath $m$}\cdot\mbox{\boldmath $\sigma$}_{\mbox{\scriptsize B}}
&\quad&
\mbox{for $|\mbox{\boldmath $m$}|=1$}.
\end{array}
\right.
\eeq

An expectation value of the outcome in the measurement on the qubit A is given by
\beq
\frac{1}{2}
\mbox{Tr}
[
E(\mbox{\boldmath $l$})
(\mbox{\boldmath $I$}_{(2)}
+
\mbox{\boldmath $a$}(\lambda)\cdot\mbox{\boldmath $\sigma$})
]
=
\frac{1}{2}
\mbox{Tr}
[(\mbox{\boldmath $l$}\cdot\mbox{\boldmath $\sigma$})
(\mbox{\boldmath $a$}\cdot\mbox{\boldmath $\sigma$})]
=
\mbox{\boldmath $l$}\cdot\mbox{\boldmath $a$}.
\lab{expectation-value-orthogonal-measurement-qubit-A}
\eeq
Equation~(\ref{expectation-value-orthogonal-measurement-qubit-A})
implies that
we obtain $1$ as an output with probability $(1+\mbox{\boldmath $l$}\cdot\mbox{\boldmath $a$})/2$
and
we
obtain $(-1)$ as an output with probability $(1-\mbox{\boldmath $l$}\cdot\mbox{\boldmath $a$})/2$
in the measurement on the qubit A.
We obtain a similar result on the qubit B.

Therefore,
we can describe an expectation value of a product of
two outputs obtained from the measurements on the qubits A and B
as follows:
\beqa
C(\mbox{\boldmath $l$},\mbox{\boldmath $m$})
&=&
\mbox{Tr}[
(E(\mbox{\boldmath $l$})_{\mbox{\scriptsize A}}
\otimes
E(\mbox{\boldmath $m$})_{\mbox{\scriptsize B}})
W(q)] \non \\
&=&
\int
d\lambda
\int^{1}_{0}
d\lambda_{\mbox{\scriptsize A}}
\int^{1}_{0}
d\lambda_{\mbox{\scriptsize B}}
\;
p(\lambda)
A(\lambda,\lambda_{\mbox{\scriptsize A}};\mbox{\boldmath $l$})
B(\lambda,\lambda_{\mbox{\scriptsize B}};\mbox{\boldmath $m$}),
\lab{W(q)-hidden-variable-model}
\eeqa
where
\beq
A(\lambda,\lambda_{\mbox{\scriptsize A}};\mbox{\boldmath $l$})
=
\left\{
\begin{array}{lll}
1 & \quad
& \mbox{for $0\leq\lambda_{\mbox{\scriptsize A}}\leq(1+\mbox{\boldmath $l$}\cdot\mbox{\boldmath $a$})/2$},\\
-1 & \quad
& \mbox{for $(1+\mbox{\boldmath $l$}\cdot\mbox{\boldmath $a$})/2<\lambda_{\mbox{\scriptsize A}}\leq 1$},
\end{array}
\right.
\eeq
and
\beq
B(\lambda,\lambda_{\mbox{\scriptsize B}};\mbox{\boldmath $m$})
=
\left\{
\begin{array}{lll}
1 & \quad
& \mbox{for $0\leq\lambda_{\mbox{\scriptsize B}}\leq(1+\mbox{\boldmath $m$}\cdot\mbox{\boldmath $b$})/2$},\\
-1 & \quad
& \mbox{for $(1+\mbox{\boldmath $m$}\cdot\mbox{\boldmath $b$})/2<\lambda_{\mbox{\scriptsize B}}\leq 1$}.
\end{array}
\right.
\eeq
This is a hidden variable model.

\section{Decomposition of the separable Werner state}
\lab{section-derivation-hidden-variable-model-Werner-state}
In this section,
we derive $p(\lambda)$, $\mbox{\boldmath $a$}(\lambda)$, and $\mbox{\boldmath $b$}(\lambda)$
given in Eq.~(\ref{W(q)-product-state-representation-continuous-lambda})
explicitly.
First, we examine the expectation value of the output
that is obtained by the measurement of $E(\mbox{\boldmath $l$})_{\mbox{\scriptsize A}}$
on the qubit A of $W(q)$.
At first we calculate this expectation value from Eq.~(\ref{definition-Werner-state}),
and then we calculate it from Eq.~(\ref{W(q)-product-state-representation-continuous-lambda}).
Next, we compare these two results.

From Eq.~(\ref{definition-Werner-state}), we obtain
\beq
\mbox{Tr}
[
(E(\mbox{\boldmath $l$})_{\mbox{\scriptsize A}}\otimes\mbox{\boldmath $I$}_{\mbox{\scriptsize B}})
W(q)
]
=
q
\bra\Psi^{-}|
E(\mbox{\boldmath $l$})_{\mbox{\scriptsize A}}\otimes\mbox{\boldmath $I$}_{\mbox{\scriptsize B}}
|\Psi^{-}\ket
=0.
\lab{measurement-qubit-A-quantum-mechanics}
\eeq
On the other hand, from Eq.~(\ref{W(q)-product-state-representation-continuous-lambda}), we obtain
\beq
\mbox{Tr}
[
(E(\mbox{\boldmath $l$})_{\mbox{\scriptsize A}}\otimes\mbox{\boldmath $I$}_{\mbox{\scriptsize B}})
W(q)
]
=
\int
d\lambda\;
p(\lambda)
(\mbox{\boldmath $l$}\cdot\mbox{\boldmath $a$}(\lambda)).
\lab{measurement-qubit-A-hidden-variable-model}
\eeq
Comparing Eqs.~(\ref{measurement-qubit-A-quantum-mechanics}) and
(\ref{measurement-qubit-A-hidden-variable-model}),
we obtain the following relation:
\beq
\int
d\lambda\;
p(\lambda)
(\mbox{\boldmath $l$}\cdot\mbox{\boldmath $a$}(\lambda))
=0
\quad\quad
\forall|\mbox{\boldmath $l$}|\leq 1.
\eeq
Thus, we arrive at
\beq
\int
d\lambda\;
p(\lambda)
a_{i}(\lambda)
=0
\quad\quad
i\in\{x,y,z\}.
\lab{qubit-A-condition}
\eeq
Examining the orthogonal measurement performed on the qubit B of $W(q)$,
we can give a similar discussion and we obtain the following result:
\beq
\int
d\lambda\;
p(\lambda)
b_{i}(\lambda)
=0
\quad\quad
i\in\{x,y,z\}.
\lab{qubit-B-condition}
\eeq

Second,
we examine the expectation value of the product of the outputs
that we obtain by the measurements of
$E(\mbox{\boldmath $l$})_{\mbox{\scriptsize A}}$
and
$E(\mbox{\boldmath $m$})_{\mbox{\scriptsize B}}$
on the qubits A and B of $W(q)$, respectively.
At first we calculate this expectation value from Eq.~(\ref{definition-Werner-state}),
and then we calculate it from Eq.~(\ref{W(q)-product-state-representation-continuous-lambda}).
Next, we compare these two results.
From Eq.~(\ref{definition-Werner-state}), we obtain
\beq
\mbox{Tr}
[
(E(\mbox{\boldmath $l$})_{\mbox{\scriptsize A}}\otimes E(\mbox{\boldmath $m$})_{\mbox{\scriptsize B}})
W(q)
]
=
q
\bra\Psi^{-}|
E(\mbox{\boldmath $l$})_{\mbox{\scriptsize A}}\otimes E(\mbox{\boldmath $m$})_{\mbox{\scriptsize B}}
|\Psi^{-}\ket
=-q(\mbox{\boldmath $l$}\cdot\mbox{\boldmath $m$}).
\lab{measurement-qubits-AB-quantum-mechanics}
\eeq
On the other hand, from Eq.~(\ref{W(q)-product-state-representation-continuous-lambda}),
we obtain
\beq
\mbox{Tr}
[
(E(\mbox{\boldmath $l$})_{\mbox{\scriptsize A}}\otimes E(\mbox{\boldmath $m$})_{\mbox{\scriptsize B}})
W(q)
]
=
\int
d\lambda\;
p(\lambda)
(\mbox{\boldmath $l$}\cdot\mbox{\boldmath $a$}(\lambda))
(\mbox{\boldmath $m$}\cdot\mbox{\boldmath $b$}(\lambda)).
\lab{measurement-qubits-AB-hidden-variable-model}
\eeq
Comparing Eqs.~(\ref{measurement-qubits-AB-quantum-mechanics})
and (\ref{measurement-qubits-AB-hidden-variable-model}),
we obtain the following relation:
\beq
\int
d\lambda\;
p(\lambda)
(\mbox{\boldmath $l$}\cdot\mbox{\boldmath $a$}(\lambda))
(\mbox{\boldmath $m$}\cdot\mbox{\boldmath $b$}(\lambda))
=
-q(\mbox{\boldmath $l$}\cdot\mbox{\boldmath $m$})
\quad\quad
\forall|\mbox{\boldmath $l$}|\leq 1,
\forall|\mbox{\boldmath $m$}|\leq 1.
\eeq
Thus, we arrive at
\beq
\int
d\lambda\;
p(\lambda)
a_{i}(\lambda)
b_{j}(\lambda)
=
-q\delta_{ij}
\quad\quad
i,j\in\{x,y,z\}.
\lab{qubits-AB-condition}
\eeq

Now, we obtain the conditions,
Eqs.~(\ref{probability-preserving-condition}),
(\ref{norm-condition}),
(\ref{qubit-A-condition}),
(\ref{qubit-B-condition}),
and
(\ref{qubits-AB-condition}),
which
$p(\lambda)$, $\mbox{\boldmath $a$}(\lambda)$, and $\mbox{\boldmath $b$}(\lambda)$
have to satisfy.
Thus,
these equations are necessary conditions for
$p(\lambda)$, $\mbox{\boldmath $a$}(\lambda)$, and $\mbox{\boldmath $b$}(\lambda)$.
However, at the same time,
they are sufficient conditions for
$p(\lambda)$, $\mbox{\boldmath $a$}(\lambda)$, and $\mbox{\boldmath $b$}(\lambda)$.
In fact,
from Eqs.~(\ref{probability-preserving-condition}),
(\ref{norm-condition}),
(\ref{qubit-A-condition}),
(\ref{qubit-B-condition}),
and
(\ref{qubits-AB-condition}),
we can always regenerate $W(q)$ defined
in Eqs.~(\ref{definition-Werner-state}) and (\ref{matrix-representation-Wq}).
For example, using
Eqs.~(\ref{W(q)-product-state-representation-continuous-lambda}),
(\ref{probability-preserving-condition}),
(\ref{qubit-A-condition}),
(\ref{qubit-B-condition}),
and
(\ref{qubits-AB-condition}),
we can calculate
the matrix element $\bra 00|W(q)|00\ket$ as follows:
\beq
\bra 00|W(q)|00\ket
=
\frac{1}{4}
\int
d\lambda\;
p(\lambda)
(1+a_{z}(\lambda))(1+b_{z}(\lambda))
=
\frac{1-q}{4}.
\eeq
This result coincides with Eq.~(\ref{matrix-representation-Wq}).
We can obtain the similar results about the other matrix elements of $W(q)$.

If we define $p(\lambda)$, $\mbox{\boldmath $a$}(\lambda)$, and $\mbox{\boldmath $b$}(\lambda)$
as described below,
they satisfy all of the necessary and sufficient conditions,
Eqs.~(\ref{probability-preserving-condition}),
(\ref{norm-condition}),
(\ref{qubit-A-condition}),
(\ref{qubit-B-condition}),
and
(\ref{qubits-AB-condition}).
First, we define the variable $\lambda$ as $\theta\in[0,\pi]$ and $\phi\in[0,2\pi)$.
Second, we define the normalized probability distribution as
\beq
p(\theta,\phi)=\frac{1}{4\pi}.
\eeq
Then we obtain
\beq
\int^{\pi}_{0}d\theta
\int^{2\pi}_{0}d\phi
\;
\sin\theta
\;
p(\theta,\phi)
=1,
\lab{probability-preserving-condition-theta-phi}
\eeq
and Eq.~(\ref{probability-preserving-condition}) is satisfied.
Here, we pay attention to the fact that
the volume element for the integral is given by $(d\theta d\phi\;\sin\theta)$
in Eq.~(\ref{probability-preserving-condition-theta-phi}).

Third,
we define $\mbox{\boldmath $a$}(\theta,\phi)$ and $\mbox{\boldmath $b$}(\theta,\phi)$
as follows:
\beq
a_{i}(\theta,\phi)=\sqrt{3q}f_{i}(\theta,\phi),
\quad\quad
b_{i}(\theta,\phi)=-\sqrt{3q}f_{i}(\theta,\phi),
\lab{definition-ab-theta-phi}
\eeq
where
\beq
\left\{
\begin{array}{lll}
f_{x}(\theta,\phi) & = & \sin\theta\cos\phi, \\
f_{y}(\theta,\phi) & = & \sin\theta\sin\phi, \\
f_{z}(\theta,\phi) & = & \cos\theta.
\lab{definition-f-xyz-theta-phi}
\end{array}
\right.
\eeq
The functions $f_{x}$, $f_{y}$, and $f_{z}$
satisfy the following relations:
\beq
\frac{1}{4\pi}
\int^{\pi}_{0}d\theta
\int^{2\pi}_{0}d\phi
\;
\sin\theta
\;
f_{i}(\theta,\phi)
=
0
\quad\quad
i\in\{x,y,z\},
\eeq
\beq
\frac{1}{4\pi}
\int^{\pi}_{0}d\theta
\int^{2\pi}_{0}d\phi
\;
\sin\theta
\;
f_{i}(\theta,\phi)f_{j}(\theta,\phi)
=
\frac{1}{3}\delta_{ij}
\quad\quad
i,j\in\{x,y,z\}.
\eeq
From the above relations,
we can confirm that
$\mbox{\boldmath $a$}(\theta,\phi)$ and $\mbox{\boldmath $b$}(\theta,\phi)$
satisfy Eqs.~(\ref{qubit-A-condition}),
(\ref{qubit-B-condition}),
and (\ref{qubits-AB-condition}).

Here, let us calculate norms of $\mbox{\boldmath $a$}$ and $\mbox{\boldmath $b$}$
from Eqs.~(\ref{definition-ab-theta-phi}) and (\ref{definition-f-xyz-theta-phi}),
\beq
|\mbox{\boldmath $a$}|
=
|\mbox{\boldmath $b$}|
=
\sqrt{3q}.
\lab{norms-ab}
\eeq
Remembering Eq.~(\ref{norm-condition}),
we can obtain a condition $0\leq q\leq 1/3$ from Eq.~(\ref{norms-ab}).
This implies that
the explicit convex combination of product states of $W(q)$ given in this section
is right only for $0\leq q\leq 1/3$.
This fact coincides with the condition for separability of $W(q)$.
From this observation, we understand
that the critical point $q=1/3$ comes from
positivity of local density operators,
$\rho_{\mbox{\scriptsize A}}(\lambda)$
and
$\rho_{\mbox{\scriptsize B}}(\lambda)$.

\section{Discussions}
\lab{section-discussions}
In this paper, we write down the separable Werner state
as a convex combination of product states explicitly,
so that we construct a hidden variable model for the separable Werner state.
Our convex combination for the separable Werner state is different
from the convex combination obtained by Wootters' method.

In our decomposition,
as shown in Eq.~(\ref{definition-ab-theta-phi}),
$\mbox{\boldmath $a$}(\lambda)$ and $\mbox{\boldmath $b$}(\lambda)$ always point
to the opposite directions with each other,
namely,
$\mbox{\boldmath $a$}(\lambda)=-\mbox{\boldmath $b$}(\lambda)$
$\forall\lambda$.
We cannot find any physical or geometrical meaning of this relation.
We are not sure whether or not
there exists a decomposed form that satisfies
$\mbox{\boldmath $a$}(\lambda)\neq -\mbox{\boldmath $b$}(\lambda)$
for some $\lambda$.

\bigskip
\noindent
{\bf \large Acknowledgment}
\smallskip

\noindent
H. A. thanks Osamu Hirota for encouragement.

\appendix
\section{The decomposed form obtained by Wootters' \\
method for the separable Werner state}
\lab{appendix-wootters-decomposition}
According to Ref.~\cite{Wootters} written by Wootters,
we can obtain the following convex combination of product states for the separable Werner state
defined in Eq.~(\ref{definition-Werner-state}) with
$0\leq q\leq 1/3$:
\beq
W(q)=\sum_{i=1}^{4}|z_{i}\ket\bra z_{i}|,
\eeq
where
\beqa
|z_{1}\ket
&=&
(1/2)
(e^{i\theta_{1}}|x_{1}\ket
+
e^{i\theta_{2}}|x_{2}\ket
+
e^{i\theta_{3}}|x_{3}\ket
+
e^{i\theta_{4}}|x_{4}\ket
), \non \\
|z_{2}\ket
&=&
(1/2)
(e^{i\theta_{1}}|x_{1}\ket
+
e^{i\theta_{2}}|x_{2}\ket
-
e^{i\theta_{3}}|x_{3}\ket
-
e^{i\theta_{4}}|x_{4}\ket
), \non \\
|z_{3}\ket
&=&
(1/2)
(e^{i\theta_{1}}|x_{1}\ket
-
e^{i\theta_{2}}|x_{2}\ket
+
e^{i\theta_{3}}|x_{3}\ket
-
e^{i\theta_{4}}|x_{4}\ket
), \non \\
|z_{4}\ket
&=&
(1/2)
(e^{i\theta_{1}}|x_{1}\ket
-
e^{i\theta_{2}}|x_{2}\ket
-
e^{i\theta_{3}}|x_{3}\ket
+
e^{i\theta_{4}}|x_{4}\ket
),
\lab{Wootters-z-basis}
\eeqa
\beqa
|x_{1}\ket
&=&
-i
\frac{\sqrt{1+3q}}{2}|\Psi^{-}\ket, \non \\
|x_{2}\ket
&=&
\frac{\sqrt{1-q}}{2}|\Psi^{+}\ket, \non \\
|x_{3}\ket
&=&
\frac{\sqrt{1-q}}{2}|\Phi^{-}\ket, \non \\
|x_{4}\ket
&=&
-i
\frac{\sqrt{1-q}}{2}|\Phi^{+}\ket,
\lab{Wootters-x-basis}
\eeqa
\beqa
|\Psi^{\pm}\ket
&=&
\frac{1}{\sqrt{2}}
(
|0\ket_{\mbox{\scriptsize A}}|1\ket_{\mbox{\scriptsize B}}
\pm
|1\ket_{\mbox{\scriptsize A}}|0\ket_{\mbox{\scriptsize B}}
), \non \\
|\Phi^{\pm}\ket
&=&
\frac{1}{\sqrt{2}}
(
|0\ket_{\mbox{\scriptsize A}}|0\ket_{\mbox{\scriptsize B}}
\pm
|1\ket_{\mbox{\scriptsize A}}|1\ket_{\mbox{\scriptsize B}}
),
\lab{Wootters-Bell-basis}
\eeqa
and
\beq
e^{-2i\theta_{1}}(1+3q)
+
(e^{-2i\theta_{2}}+e^{-2i\theta_{3}}+e^{-2i\theta_{4}})(1-q)=0.
\lab{Wootters-decomposition-condition}
\eeq
Equation~(\ref{Wootters-decomposition-condition}) does not determine
$\theta_{1}$, $\theta_{2}$, $\theta_{3}$, and $\theta_{4}$ uniquely.
For example, we have a special solution,
$\theta_{1}=0$,
$\theta_{2}=\pi/2$,
\beq
\cos\theta_{3}=\sqrt{\frac{1-3q}{2(1-q)}},
\quad\quad
\sin\theta_{3}=\sqrt{\frac{1+q}{2(1-q)}},
\eeq
and
\beq
\cos\theta_{4}=-\sqrt{\frac{1-3q}{2(1-q)}},
\quad\quad
\sin\theta_{4}=\sqrt{\frac{1+q}{2(1-q)}}.
\eeq
By substituting the above special solution into
Eqs.~(\ref{Wootters-z-basis}), (\ref{Wootters-x-basis}),
and (\ref{Wootters-Bell-basis}),
we can confirm that
$|z_{1}\ket$, $|z_{2}\ket$, $|z_{3}\ket$, and $|z_{4}\ket$
are product states.

\end{document}